# Phenomenology of Goldstino Couplings

LU-XIN LIU

*Department of Physics, Purdue University,*
*West Lafayette, IN 47907, USA*
liul@physics.purdue.edu

A general coupling of the Goldstino to the matter field and the weak gravitational field is constructed based on the standard and the nonlinear Volkov-Akulov realization of SUSY. The resulting Lagrangian, which is invariant under SUSY transformations, can give rise to explicit interactions which couple the helicity $\pm\frac{1}{2}$ states of the gravitino with the gravitational field as well as the matter field.

The realization of local SUSY leads to the supergravity theory, from which the irreducible representation of supersymmetry is carried by the spin-2 graviton and its spin-$\frac{3}{2}$ partner, the gravitino.[1–4] SUSY can also be realized nonlinearly from a massless Weyl spinor field constructed by Volkov and Akulov.[5] Here, the Lagrangian directly gives rise to spontaneous supersymmetry breaking. When it couples to supergravity, the resulting local invariance gives an effective mass to the spin-$\frac{3}{2}$ component of supergravity after the spontaneous broken supersymmetry is gauged.[6–7] As a result, through the super Higgs mechanism, the spin -$\frac{1}{2}$ spinor, behaving as the Nambu-Goldstone fermion, provides the helicity $\pm\frac{1}{2}$ degrees of freedom of the spin -$\frac{3}{2}$ gravitino.

A superfield formulation of the nonlinear realization of SUSY was developed.[8,9,15] Recently, however, an alternative formulation of supersymmetric invariant couplings of the Goldstino to matter was constructed.[10–11] Using the standard realization of supersymmetry, the SUSY algebra can be nonlinearly realized on the matter field which is free to carry an arbitrary set of Lorentz or internal symmetry indices. In addition, through the special properties of the Akulov-Volkov Lagrangian, the whole action containing the Goldstino and matter field is then constructed. It is invariant under both SUSY and internal symmetry transformations. This elegant procedure enables one to work directly with the ordinary Goldstino and matter field, and any operator can be made part of the supersymmetric action. Further work successfully extends this formulation to the general N-extended supersymmetry.[12] It can also be applied to the scale symmetry and superconformal symmetry.[13,14]

In this paper, based on this alternative formalism of the Goldstino couplings with the matter field, we extend Refs. 10 and 11 to include the gravitational field. Our procedure starts from the weak gravitational field assumption, followed by the construction of the nonlinearly realized supermymmetric action in the Minkowski four-dimensional space. The resulting formalism of the globally invariant action gives us the general explicit interactions that involve the matter field, the gravitation field and the Goldstino. The matter field can be any field that has Lorentz and internal indices. By this simple and straightforward method, the action is free of extra superfield or a huge amount of superstructures.

Another purpose of this paper is to get physical processes that govern certain light gravitino interactions with the matter field and the graviton by using the supersymmetric equivalence theorem.[17] The equivalence theorem holds in general in theories which possess gauge properties spontaneously broken by a Higgs-type mechanism. When the energy scale is much greater than $m_{3/2}$, the amplitude of the interaction with the longitudinal gravitino could be equivalent to amplitudes with corresponding external Goldstinos. Consequently, from our action, we obtain some of the most important consequences of the supergravity theory which govern the process of certain light gravitino (helicity $\pm\frac{1}{2}$) that interacts with the matter field and the graviton.

As a phenomenological consequence, the additional channel of the gravitino cooling of supernova is presented here in the process of the supernova explosion. As a result, one may need to consider the contribution from this viewpoint while estimating the lower bound on the decay constant $F$. It is also pointed out that we can complete the SUSY invariant action by including the standard model providing that we redefine the gauge covariant field strength tensor and its SUSY transformation.

The nonlinear realized SUSY was introduced by Akulov and Volkov in 1972.[5] The transformation of the Goldstino Weyl spinor field $\lambda$ is

$$\delta_\xi \lambda^\alpha = F\xi^\alpha - i\frac{1}{F}(\lambda\sigma^\rho \bar{\xi} - \xi\sigma^\rho \bar{\lambda})\partial_\rho \lambda^\alpha , \quad (1)$$

$$\delta_\xi \bar{\lambda}_{\dot{\alpha}} = F\bar{\xi}_{\dot{\alpha}} - i\frac{1}{F}(\lambda\sigma^\rho \bar{\xi} - \xi\sigma^\rho \bar{\lambda})\partial_\rho \bar{\lambda}_{\dot{\alpha}} , \quad (2)$$

where $\xi, \bar{\xi}$ are Weyl spinor SUSY transformation parameters, and $F$ is the decay constant of the Goldstino field. The SUSY algebra is closed as

$$(\delta_\eta \delta_\xi - \delta_\xi \delta_\eta)\lambda^\alpha = -2i(\eta\sigma^\rho \bar{\xi} - \xi\sigma^\rho \bar{\eta})\partial_\rho \lambda^\alpha ; \quad (3)$$

the SUSY invariant Lagrangian which describes the massless Goldstino self-dynamics is

$$\ell_{AV} = -\frac{F^2}{2}\det A , \quad (4)$$

where the Volkov-Akulov vierbein takes the form

$$A_m^a = \delta_m^a - i\frac{1}{F^2}\partial_m \lambda \sigma^a \bar{\lambda} + i\frac{1}{F^2}\lambda\sigma^a \partial_m \bar{\lambda} , \quad (5)$$

then the Akulov-Volkov Lagrangian transforms as a total spacetime divergence

$$\delta(\xi,\bar{\xi})\ell_{AV} = -i\frac{1}{F}\partial_\alpha[(\lambda\sigma^\alpha\bar{\xi} - \xi\sigma^\alpha\bar{\lambda})\ell_{AV}].$$

Therefore the action

$$I_{AV} = \int d^4x\,\ell_{AV} = -\frac{F^2}{2}\int d^4x\,\det A \tag{6}$$

is SUSY invariant

$$\delta(\xi,\bar{\xi})I_{AV} = 0.$$

The Akulov-Volkov Lagrangian can be expanded as

$$\ell_{AV} = -\frac{F^2}{2} - \frac{i}{2}(\lambda\sigma^m\partial_m\bar{\lambda} - \partial_m\lambda\sigma^m\bar{\lambda}) + \text{other terms} \tag{7}$$

it is explicit to see that this Lagrangian has a nonzero vacuum expectation value, and it directly gives rise to the spontaneous supersymmetric breaking. The SUSY algebra can also be realized nonlinearly based on the standard realization of the matter field[15] $\phi_i$

$$\delta(\xi,\bar{\xi})\phi_i = -i\frac{1}{F}(\lambda\sigma^\rho\bar{\xi} - \xi\sigma^\rho\bar{\lambda})\partial_\rho\phi_i, \tag{8}$$

where the index $i$ represents any Lorentz or internal symmetry labels. Using the Akulov-Volkov Lagrangian, we may construct the SUSY invariant action which contains both the Goldstino and the matter fields. Its generic form can be written as[10,11]

$$I_{eff} = \int d^4x\,\det A\,\ell_\phi(\phi,\tilde{D}\phi), \tag{9}$$

where the nonlinearly realized SUSY covariant derivative is defined as

$$\tilde{D}_\mu = (A^{-1})_\mu^\nu\partial_\nu, \tag{10}$$

and $(A^{-1})_\mu^\nu A_\nu^\rho = \delta_\mu^\rho$. Therefore, according to the standard realization of SUSY, it transforms as

$$\delta(\xi,\bar{\xi})\tilde{D}_\mu\phi_i = -i\frac{1}{F}(\lambda\sigma^\rho\bar{\xi} - \xi\sigma^\rho\bar{\lambda})\partial_\rho(\tilde{D}_\mu\phi_i), \tag{11}$$

from which it follows that the action

$$I_{LL} = -2\frac{1}{F^2}\int d^4x\,\ell_\phi(\phi,\tilde{D}\phi)\ell_{AV}(\lambda,\bar{\lambda}) \tag{12}$$

is invariant under the nonlinear SUSY transformation, i.e.

$$\delta(\xi,\bar{\xi})I_{LL} \propto \int\{\delta(\xi,\bar{\xi})\det A\}\ell_\phi + \det A\{-i\frac{1}{F}(\lambda\sigma^\rho\bar{\xi} - \xi\sigma^\rho\bar{\lambda})\partial_\rho\ell_\phi\}$$

$$\propto \int\{-i\frac{1}{F}(\lambda\sigma^\rho\bar{\xi} - \xi\sigma^\rho\bar{\lambda})\}\partial_\rho(\det A\,\ell_\phi)$$

$$= 0$$

We may observe that, in the absence of Goldstino fields, this action reduces to the ordinary matter action $\int d^4x\,\ell_\phi(\phi,\partial_\mu\phi)$. So the action $I_{LL}$ contains the ordinary matter action as well as the couplings of the Goldstino to matter.

For the case when the gravitational field is present, we focus on the particle aspect of the gravity instead of the standard geometric approach based on Einstein's GR. The metric tensor field $g_{\mu\nu}$ is expanded around some classical solution $\eta_{\mu\nu}$, i.e. $g_{\mu\nu} = \eta_{\mu\nu} + 2kh_{\mu\nu}$,[16] where $\eta_{\mu\nu}$ describes the constant structure background, $h_{\mu\nu}$ is the dynamic part which describes the gravitational field and $k^2 = 8\pi G_N$, where $G_N$ is the Newtonian gravitational constant. When the gravitational field is weak, we could work on the linear Einstein gravity directly. Also, we do not need to introduce the auxiliary fields and the super structures. The Lagrangian which describes the massless spin-2 field is

$$\ell_G = \frac{1}{2k}(R_{\mu\nu} - \frac{1}{2}\eta_{\mu\nu}R)h^{\mu\nu}. \tag{13}$$

The linearlized Ricci tensor is given by

$$R_{\mu\nu} = -k(-\partial^\alpha \partial_\alpha h_{\mu\nu} + \partial_\lambda \partial_\mu h^\lambda_\nu + \partial_\nu \partial_\lambda h^\lambda_\mu - \partial_\mu \partial_\nu h^\lambda_\lambda)$$

and we have the original gauge transformation of $h_{\mu\nu}$

$$\delta^g h_{\mu\nu} = \frac{1}{2k}[\partial_\mu \varsigma_\nu(x) + \partial_\nu \varsigma_\mu(x)] . \tag{14}$$

The field $h_{\mu\nu}$ has the Goldstino dependent transformation

$$\delta(\xi,\bar\xi)h_{\mu\nu} = -i\frac{1}{F}(\lambda\sigma^\rho \bar\xi - \xi\sigma^\rho \bar\lambda)\partial_\rho h_{\mu\nu}$$

and

$$\delta(\tilde D_\alpha h_{\mu\nu}) = -i\frac{1}{F}(\lambda\sigma^\rho \bar\xi - \xi\sigma^\rho \bar\lambda)\partial_\rho(\tilde D_\alpha h_{\mu\nu}),$$

where $\tilde D_\mu = (A^{-1})^\nu_\mu \partial_\nu$. We further require the gauge transformation parameter $\varsigma_\mu(x)$ to transform under the standard realization of the supersymmetry. Therefore

$$\delta(\xi,\bar\xi)\varsigma_\mu(x) = -i\frac{1}{F}(\lambda\sigma^\rho \bar\xi - \xi\sigma^\rho \bar\lambda)\partial_\rho \varsigma_\mu(x). \tag{15}$$

This is the SUSY transformation property of $\varsigma_\mu(x)$, basically different from its ordinary gauge transformation. Just like the gauge and supersymmetric transformation in the N=1 super-QED model, we can form a closed algebra when imposing the gauge and supersymmetric variation on the same field $h_{\mu\nu}$, then the gauge and SUSY transformation commute:

$$[\delta^g, \delta(\xi,\bar\xi)]h_{\mu\nu} = \delta^g(-i\frac{1}{F})(\lambda\sigma^\rho \bar\xi - \xi\sigma^\rho \bar\lambda)\partial_\rho h_{\mu\nu} - \delta(\xi,\bar\xi)\frac{1}{2k}[\partial_\mu \varsigma_\nu(x) + \partial_\nu \varsigma_\mu(x)]$$

$$= (-i\frac{1}{F})(\lambda\sigma^\rho \bar\xi - \xi\sigma^\rho \bar\lambda)\partial_\rho\{\frac{1}{2k}[\partial_\mu \varsigma_\nu(x) + \partial_\nu \varsigma_\mu(x)]\}$$

$$- \frac{1}{2k}\delta(\xi,\bar\xi)[\partial_\mu \varsigma_\nu(x) + \partial_\nu \varsigma_\mu(x)] = 0.$$

Hence we get the SUSY invariant action after replacing all the ordinary derivative by the covariant derivative $\tilde D_\mu = (A^{-1})^\nu_\mu \partial_\nu$,

$$I_G = -2\frac{1}{F^2}\int d^4x \ell_{AV} \ell_G. \tag{16}$$

Now we can introduce the action to contain the Goldstino and the matter fields as well as the gravitational field. By replacing all the space-time derivative by the nonlinearly realized SUSY covariant derivative, and comparing with Ref.16, the total action of the whole system is constructed as

$$I = I_{AV} + I_{LL} + I_G + I_{\lambda Q} + I_{LG}. \tag{17}$$

It is SUSY invariant

$$\delta(\xi, \bar{\xi})I = 0,$$

where $I_{AV}$ is the action of the Goldstino field, $I_{LL}$ is the normal action of matter coupling with Goldstino given by Eq.(12), and $I_{\lambda Q}$ is the SUSY current interaction [10]

$$I_{\lambda Q} = -2\frac{1}{F^2} k \int d^4 x \ell_{AV} (\tilde{D}_\mu \lambda^\alpha Q^\mu_{\phi\alpha} + \overline{Q}^\mu_{\phi\dot\alpha} \tilde{D}_\mu \bar\lambda^{\dot\alpha}).$$

The associated supersymmetry current $Q^\mu_{\phi\alpha}$ and $\overline{Q}^\mu_{\phi\dot\alpha}$, which are constructed from pure matter actions for linear realizations of SUSY, have the same transformation law as Eq.(8) with all the spacetime derivative replaced by nonlinearly realized SUSY covariant derivatives. The $I_{LG}$ term is the interaction with the spin-2 field,

$$I_{LG} = -2k\frac{1}{F^2} \int d^4 x \ell_{AV} T^{\mu\nu} h_{\mu\nu}. \tag{18}$$

Here, $T^{\mu\nu}$ is the energy-momentum tensor and once again with all spacetime derivative replaced by the covariant derivative. For a light gravitino, we only focus on the interactions of the helicity $\pm\frac{1}{2}$ modes which are enhanced as a result of the supersymmetric equivalence theorem. On dimension ground, it is obvious that those modes which interact only with the matter field or the graviton are contained in $I_{LL}, I_G$ and $I_{LG}$. The gravitational Lagrangian $\ell'_G = -2\frac{1}{F^2}\ell_{AV}\ell_G$ can be expanded in orders of $F$. We consider the first two lowest order terms

$$\ell'_G = -2\frac{1}{F^2}\ell_{AV}\ell_G$$

$$= h^{\mu\nu} E_{\mu\nu} + i\frac{1}{F^2}(\lambda\sigma^m \partial_m \bar\lambda - \partial_m \lambda \sigma^m \bar\lambda) E_{\mu\nu} h^{\mu\nu}$$

$$+ i\frac{1}{F^2} E'_{\mu\nu} h^{\mu\nu} + i\frac{1}{F^2} E''^{\mu\nu} h_{\mu\nu} + \ldots, \tag{19}$$

where

$$E_{\mu\nu} = \frac{1}{2k}(R_{\mu\nu} - \frac{1}{2}\eta_{\mu\nu}R),$$

$$E'_{\mu\nu} = \frac{1}{2k}(R'_{\mu\nu} - \frac{1}{2}\eta_{\mu\nu}R'),$$

$$E''_{\mu\nu} = \frac{1}{2k}(R''_{\mu\nu} - \frac{1}{2}\eta_{\mu\nu}R''),$$

$$R'_{\mu\nu} = -k(-\tilde\partial^\alpha \partial_\alpha h_{\mu\nu} + \tilde\partial_\lambda \partial_\mu h^\lambda_\nu + \tilde\partial_\nu \partial_\lambda h^\lambda_\mu - \tilde\partial_\mu \partial_\nu h^\lambda_\lambda),$$

$$R''_{\mu\nu} = -k(-\partial^\alpha \tilde\partial_\alpha h_{\mu\nu} + \partial_\lambda \tilde\partial_\mu h^\lambda_\nu + \partial_\nu \tilde\partial_\lambda h^\lambda_\mu - \partial_\mu \tilde\partial_\nu h^\lambda_\lambda),$$

$$\tilde{\partial}_\mu = (\partial_\mu \lambda \sigma^\nu \bar{\lambda} - \lambda \sigma^\nu \partial_\mu \bar{\lambda}) \partial_\nu .$$

The first term gives the normal Lagrangian of the free gravitational field, while the $\frac{1}{F^2}$ part yields the interaction between the Goldstino and $h_{\mu\nu}$. It describes the interaction of two helicity $\pm \frac{1}{2}$ gravitinos with two gravitons. The last term in Eq. (17) describes the coupling of the matter field to both the gravitino and the graviton

$$\begin{aligned}
\ell'_{LG} &= -2k \frac{1}{F^2} \ell_{AV} T^{\mu\nu} h_{\mu\nu} \\
&= kT^{\mu\nu} h_{\mu\nu} + ik \frac{1}{F^2} \lambda \sigma^\beta \partial_\beta \bar{\lambda} h_{\mu\nu} T^{\mu\nu} \\
&\quad - ik \frac{1}{F^2} \partial_\beta \lambda \sigma^\beta \bar{\lambda} h_{\mu\nu} T^{\mu\nu} + \ldots .
\end{aligned} \quad (20)$$

The leading term is the normal interaction between the field $\phi$ and $h_{\mu\nu}$, and the rest are just the couplings of $\phi$ with both $h_{\mu\nu}$ and $\lambda$, which describe processes involving the emission or absorption of helicity $\pm \frac{1}{2}$ gravitino as well as the the graviton.

As an example of the important physical consequences of the SUSY invariant action in Eq.(17), we consider the gravitino cooling of the supernova. Besides the ordinary channel of gravitino pair production as pointed out in Ref.11, we note from Eq.(19) that another channel which involves the interaction of the graviton and the helicity $\pm \frac{1}{2}$ modes of the light gravitino is present. In a theory where the mass scales of gauginos and the superpartners of light fermions are above the core temperature of the supernova, this channel need to be included when we estimate the lower bound on $F$ via the supernova cooling rate. We plan to apply our analysis to a detailed investigation along this topic in a subsequent paper.

It is possible to include other higher derivative terms in the effective action of Eq.(17), and these terms become important only at higher energies.[15] When we consider the internal symmetry of the matter field, the redefined gauge field strength tensor $\tilde{F}^a_{\mu\nu} = (A^{-1})^\alpha_\mu (A^{-1})^\beta_\nu F^a_{\alpha\beta}$, where $F^a_{\alpha\beta} = \partial_\alpha A^a_\beta - \partial_\beta A^a_\alpha + if^{abc} A^b_\alpha A^c_\beta$ is the usual field strength, transforms in the same way as Eq. (8),[11]

$$\delta(\xi, \bar{\xi}) \tilde{F}^a_{\mu\nu} = -i \frac{1}{F} (\lambda \sigma^\rho \bar{\xi} - \xi \sigma^\rho \bar{\lambda}) \partial_\rho \tilde{F}^a_{\mu\nu} .$$

We can also include the internal gauge field and the gravitational force together in the same SUSY invariant Lagrangian, which has the general form

$$I = \int d^4 x \det A \ell_{eff} (\tilde{D}_\mu \lambda, \tilde{D}_\mu \bar{\lambda}, \phi^i, \tilde{D}_\mu \phi^i, \tilde{F}_{\mu\nu}, h_{\mu\nu}) , \quad (21)$$

where $\ell_{eff}$ is any gauge invariant function as in Eq.(17).

In conclusion, we have obtained the general form of the Goldstino coupling Lagrangian which includes the Goldstino, ordinary matter field, the gauge field as well as the gravitational field. When the internal symmetry is taken as $SU(3) \times SU(2) \times U(1)$, it is natural to group the standard model with the gravity in the same SUSY invariant Lagrangian, which provides a valid description of the low energy and weak gravitational interactions in M4 space. It is actually a theory where supersymmetry is spontaneously broken in a hidden sector at a much higher energy scale. The low energy effective theory constructed here communicates to the known particles of the standard model. Also, we expect interesting physical consequences directly from the general formulation of Goldstino couplings (Eq.21).

**Acknowledgments**

This paper is partially supported by the Purdue Research Foundation (PRF) in the summer of 2004. The author thanks S.T.Love for helpful discussions. The author is grateful to T.K.Kuo for carefully reading the manuscript. Also the author thanks M.Burkardt, Y.X.Gui and H.S.Song for their kindness and support.